# Speckle suppression in digital in-line holographic microscopy through liquid crystal dynamic scattering

Emilia Wdowiak,[1,2,†] Nathan Spiller,[2] Tianxin Wang,[2] Camron Nourshargh,[2] Jolanta Mierzejewska,[3] Piotr Zdańkowski,[1] Stephen M. Morris,[2] Steve J. Elston,[2] Maciej Trusiak,[1] Martin J. Booth [2,*]

[1] *Warsaw University of Technology, Institute of Micromechanics and Photonics, 8 Sw. A. Boboli St., 02-525 Warsaw, Poland*
[2] *University of Oxford, Department of Engineering Science, Parks Road, OX1 3PJ, UK*
[3] *Warsaw University of Technology, Chair of Drug and Cosmetics Biotechnology, 3 Noakowskiego St., 00-664 Warsaw, Poland*
[†] *emilia.wdowiak.dokt@pw.edu.pl*
[*] *martin.booth@eng.ox.ac.uk*

**We demonstrate speckle noise reduction in an in-line holographic imaging system using a Zwitterion-doped liquid crystal dynamic scatterer (LCDS) cell diffuser. Integrated into a minimally modified bright-field microscope, the LCDS actively modulates system's spatial coherence. The proposed solution suppresses coherent artifacts without introducing bulky moving parts, while enhancing image resolution and preserving overall system simplicity. Quantitative performance tested on a phase and amplitude test targets, as well as phase-amplitude biological sample, shows significant noise reduction and methods versatility. Though validated in a holographic in-line setup, the approach is applicable to other imaging techniques requiring compact, vibration-free speckle suppression.**

## INTRODUCTION

In-line digital holographic microscopy (DHM) relies on coherent light sources to encode complex optical field through interference [1]. While such coherence is essential for retrieving phase information, it also introduces a well-known drawback - coherent noise, commonly referred to as speckle noise - which degrades both resolution and signal-to-noise ratio (SNR) [2]. Speckle patterns in recorded interferograms/holograms propagate directly into the reconstructed amplitude and phase maps, reducing the fidelity of the recovered complex optical field.

Speckle noise originates from both temporal and spatial coherence [2]. High temporal coherence, set by the narrow spectral linewidth of the source, allows interference over large optical path differences, while high spatial coherence, resulting from the small angular extent of a quasi–point source, produces high-contrast interference fringes across the field. When such coherent light interacts with the microgeometry of the sample surface (e.g., roughness, deposited dust), multiple scattering events produce random interference patterns that manifest as speckles at the imaging plane. These artifacts are particularly problematic as they strongly depend on the object's shape and surface characteristics. Speckle noise poses a substantial challenge for traditional numerical image filtration algorithms, which generally struggle to differentiate genuine high-frequency features from noise-induced artifacts [3,4].

Various optical strategies have emerged aiming at noise minimization, such as limiting temporal coherence by employing LED illumination or filtered broadband sources [3,5]. However, such methods reduce achievable fringe density and directly truncate available spectral bandwidth. A prevalent group of speckle noise reduction techniques relies on introducing noise diversity while maintaining signal temporal coherence. These methods operate by continuously changing the phase delays in the optical path, through a medium generating different speckle patterns, enabling effective decorrelation of coherent noise. Among numerous strategies proposed, the rotating or simply moving diffusers stand out as the most widely utilized approaches due to their intuitive implementation [6–8]. However, mechanical movement naturally introduces vibrations, complicating its integration into vibration-sensitive interferometric setups, and adversely affecting the measurement stability. Alternative approaches, including varying illumination angles [9], translating sensor positions [10], using multimode optical fiber [11], spatial light modulators or piezoelectric actuators [12], also exist but generally involve complex or bulky hardware adjustments.

To overcome these limitations, we propose a novel hardware approach using liquid crystal dynamic scatterer (LCDS) cells as speckle reducer. This elegant electro-optic device provides speckle noise diversity through electrohydrodynamic instabilities (EHDIs) in the liquid crystal mixture without introducing mechanical vibrations. This makes it particularly advantageous for interferometric and holographic applications. Due to its compact size and the absence of moving parts, the LCDS cell can significantly enhance the compactness and robustness of holographic imaging systems intended for technical and biomedical measurements. When an AC electric field (1 kHz) is applied to the liquid crystal material, flow can be induced through EHDIs. At particular field amplitudes, this flow becomes turbulent, with the liquid crystal director orientation becoming spatially and temporally random. This turbulent flow regime dynamically scatters incident light, generating a series of statistically independent, decorrelated speckle patterns within short a time interval [13]. Moreover, such LCDS cells have not been previously utilized in interferometry nor holography, therefore this contribution represents a promising and innovative solution to the persistent challenge of speckle noise in coherent imaging. In this paper, we demonstrate that LCDS effectively reduces speckle noise without compromising the resolution of the imaging system. Specifically, we integrate the LCDS into a compact, recently introduced, in-line DHM system [14,15]. We illustrate significant noise reduction capabilities through experimental imaging of phase and amplitude test targets and biological sample - *Saccharomyces cerevisiae* yeast, highlighting the practical and versatile utility of our proposed LCDS approach.

## PROPOSED SETUP

The DHM method used in this work is based on a recently reported implementation of Gabor in-line holography adapted to a conventional bright-field microscope [14,15]. The key insight from that work is that complex optical field retrieval can be achieved with just two simple modifications: replacing the incoherent illumination source with a coherent one, and axially displacing either the sample or the detector from the system's focal plane. This minimal yet powerful modification transforms a standard bright-field microscope into a phase-sensitive imaging system, making it highly attractive for systems where bright-field imaging is already integrated. It is particularly well-suited for real-time preview or monitoring in complex fabrication systems, such as, tracking the refractive index changes during two-photon polymerization or monitoring other microfabrication processes, where additional imaging components must be kept as simple as possible [16–18]. In such minimally adaptable systems, physical space for hardware additions is often very constrained. Moreover, the introduction of coherent illumination, while necessary to produce the Gabor interference fringes carrying complex field information, inevitably leads to the presence of coherent (speckle) noise. This is where the use of a LCDS becomes particularly advantageous: when appropriate voltage is applied, it actively reduces spatial coherence without requiring mechanical motion, occupies minimal physical space, and preserves overall system compactness.

Figure 1 illustrates the optical system used throughout this study. The illumination source is a single-mode DPSS laser operating at a central wavelength of $\lambda = 532$ nm with narrow spectral linewidth (< 0.01 pm FWHM). To achieve speckle reduction, only two components are required: a focusing optic and the LCDS cell, which is less than 3 mm thick. A collimated laser beam is focused onto the LC layer using a condenser lens (0.25 NA, 10×), with the focal plane precisely aligned to the liquid crystal layer for optimal performance. Accurate positioning at the focal plane is crucial, as even slight defocusing can over-suppress spatial coherence, diminishing the contrast of Gabor fringes and impairing the reconstruction. In contrast to mechanically actuated speckle reducers, such as rotating diffusers, the liquid crystal-based approach is immune to misalignment issues like off-axis wobble, which would otherwise introduce unwanted spatial coherence variations.

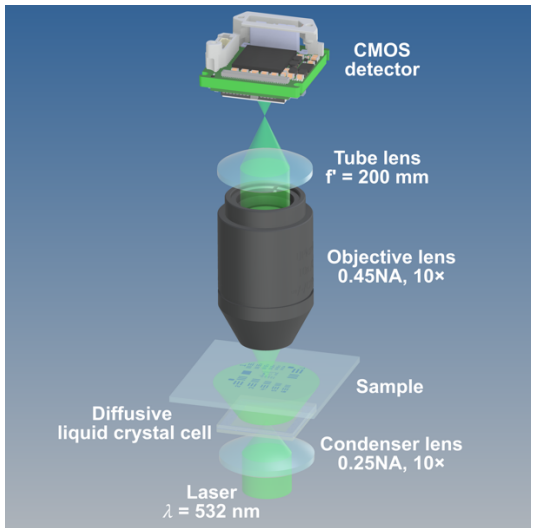

Fig. 1. Phase imaging system used in this study, including: $\lambda = 532$ nm laser source, illumination 0.25 NA, 10× condenser lens, diffusive LCDS device, sample, collecting 0.45 NA, 10× objective lens, tube lens f' = 200 mm, and CMOS detector.

The diffused light then illuminates the sample, positioned less than 1 cm from the LCDS cell, with closer placement preferred to minimize illumination falloff. Signal from the sample is collected by a 0.45 NA, 10× microscope objective. The imaging path includes a 200 mm focal length tube lens, which forms the intermediate image approximately 2 mm in front of the detector plane. The detector is a monochromatic CMOS camera (5496 × 3672 pixels, 2.4 × 2.4 µm² pixel size). In this setup, the axial offset is introduced by axially shifting the detector, rather than the sample. This decision aligns with the intended role of the system as a modular add-on for existing setup (e.g., fabrication or inspection systems), where the sample position is often fixed by the primary system design.

The detector captures a defocused image of the sample – a hologram containing a diffractive fringe pattern (Gabor fringes) that encodes the object's complex optical field information. To retrieve the information from the image plane, numerical backpropagation is required. Complex field reconstruction can be performed from a single captured frame - suitable for dynamic measurements - or from multiple frames to mitigate the twin-image artifact, a known limitation of in-line holography [6,14,15]. In this study, five frames were acquired, each positioned 0.6 mm farther from the tube lens. The first defocused frame was acquired approximately 2 mm away from the focal plane. The retrieval was carried out using the Gerchberg–Saxton iterative algorithm, incorporating field filtering constraints, with five iterations applied for convergence [15].

## DIFFUSIVE LIQUID CRYSTAL CELL

*Fabrication:*

The LCDS devices were fabricated to enable active modulation of the spatial coherence of laser light for the reduction of laser speckle and other unwanted coherence artefacts. Devices were assembled using commercially available liquid crystal cells (INSTEC S100A200uT180) with nominal cell gaps of 20 µm (±2 µm). Prior to filling, the exact cell gap thicknesses were verified by analysing interference fringes across the 300–1000 nm wavelength range using an Agilent Cary 8454 UV-Vis spectrophotometer, and only cells within 5% of the nominal gap were selected for use. The cells featured prefabricated transparent indium-tin-oxide (ITO) electrodes (23 nm thick) and an alignment layer (DPI-V011) on the inner surfaces of the substrates to promote homeotropic alignment of the liquid crystal at the boundaries. These features are intrinsic to the purchased cells and not modified as part of the device fabrication process. The active area of the electrodes defined a 10 mm × 10 mm region for the application of the electric field.

The liquid crystal mixtures consisted of a base host doped with small concentrations (<1 wt.%) of a zwitterionic additive (Reichardt's dye), which has previously been optimized for applications requiring speckle suppression [13]. The mixtures were capillary-filled into the cells at approximately 5°C above the clearing point to ensure complete isotropic flow and to minimize the risk of air bubble formation during filling. After assembly, uniform dispersion of the liquid crystal within the cell was confirmed using a polarizing optical microscope (Olympus BX51-P).

*Operation:*

Upon application of an alternating electric field across the LCDS, the liquid crystal mixture undergoes a transition from a uniform homeotropic alignment to a dynamic scattering mode (DSM) driven by EHDIs. These instabilities arise from the interplay between the material's dielectric anisotropy and ionic conductivity, which generate competing torques on the liquid crystal director [19]. As the field amplitude exceeds the threshold for instability, flow patterns form within the liquid crystal layer, disrupting the uniform molecular orientation and giving rise to transient, spatially varying refractive index fluctuations. This turbulent director reorientation leads to a dynamic randomization of the optical phase of light propagating through the cell. For incident coherent light, the presence of temporally evolving scattering domains results in the generation of a sequence of partially decorrelated wavefronts, which in turn suppress the magnitude of coherence-based artefacts such as speckle noise. The extent of speckle reduction can be precisely tuned by adjusting the amplitude and frequency of the applied electric field, allowing for application-specific optimization of speckle suppression.

For this study the LCDS was operated with a zero-mean 400 $V_{PP}$ square wave at 1 kHz.

## RESULTS AND DISCUSSION

To evaluate the performance of the LCDS device and its impact on system resolution, phase and amplitude USAF tests charts were measured (Fig. 2). To quantify speckle reduction performance, the standard deviation (*std*) was calculated within a background region free of sample structure (where the ideal *std* is zero) [20]. Figure 2 features the areas of *std* calculation with white dotted frames.

The left column of Fig. 2 displays individual holograms selected from a set of five, captured for each case: a phase test target without (Fig. 2(a1)) and with LCDS speckle suppression (Fig. 2(b1)), and an amplitude target without (Fig. 2(c1)) and with suppression (Fig. 2(d1)). Correspondingly, the right column presents the reconstructed phase maps (Figs. 2(a2), 2(b2)) and amplitude maps (Figs. 2(c2), 2(d2)) for each condition. The LCDS cell was completely removed from the optical setup for measurements performed without speckle suppression. Each image includes enlarged regions highlighting the smallest resolvable features: 2 µm bars in the phase test and 0.775 µm bars in the amplitude test. In the phase reconstructions, the 2 µm resolution remains preserved with and without LCDS speckle reduction (the feature size limited by the absence of bars smaller than 2 µm on the available phase target). In the amplitude, however, denoised reconstructions show the enhancement of resolution - the 0.775 µm bars, which are blurred beyond recognition in the noisy case - Fig. 2(c2), become distinctly visible in Fig. 2(d2). In both modalities, speckle noise distorts the reconstructed bar shapes, while LCDS use restores their clarity and contrast. It is illustrated by the cross-sections in Figs. 2(d2) and 2(c2), where the amplitude map is visibly corrupted by coherent noise. We can assume that the phase resolution performs similarly when imaging details smaller than 2 µm, since the denoised hologram (Fig. 2(b1)) features well-defined dense Gabor fringes despite the reduced spatial coherence introduced by the LCDS. A similar fringe pattern is visible in Fig. 2(a1), though fringes are significantly degraded by strong speckle noise.

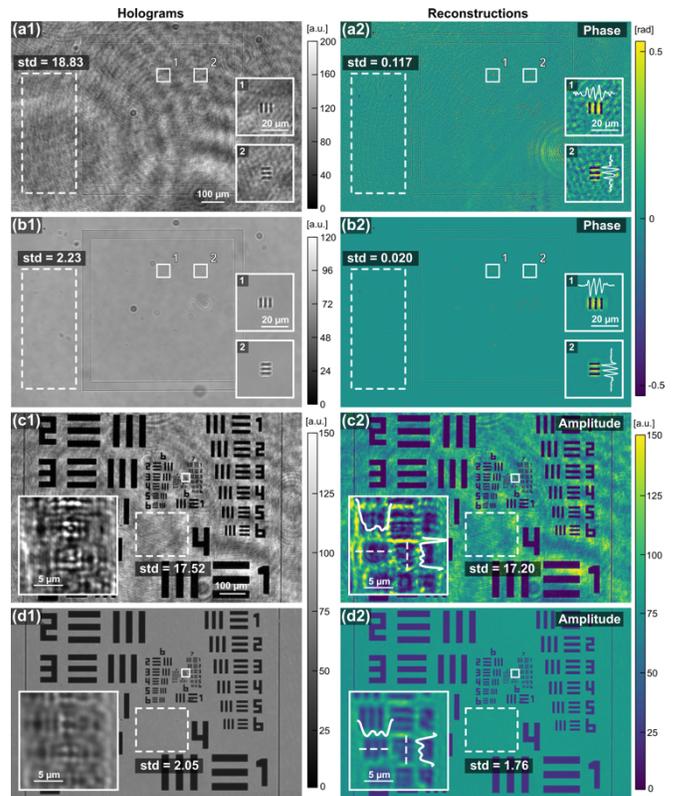

Fig. 2. Evaluation of the LCDS speckle reduction on a phase and amplitude resolution test targets. (a1, b1, c1, d1) Raw holograms acquired without (a1, c1) and with (b1, d1) the LCDS. (a2, b2) Corresponding reconstructed phase maps. Enlarged regions marked with 1 and 2 highlight Group 9 features (2 µm bars). (c2, c2) Corresponding reconstructed amplitude maps. Enlarged regions highlight Group 9 Element 3 features (0.775 µm bars).

A clear speckle reduction is observed both visually and quantitatively. Standard deviation measurements reveal an almost eightfold decrease of the *std* in both phase and amplitude holograms. In the reconstructed phase maps, the improvement reaches nearly sixfold, while in the amplitude reconstructions - where speckles have the greatest impact - the reduction is about tenfold. This substantial improvement confirms the effectiveness of the LCDS device in enabling complex optical field measurements in in-line DHM without any loss of resolution.

LCDS provokes the illumination beam intensity drop (see Supplementary Video 1), therefore the exposure time has to be adjusted when the active LCDS diffuser is introduced to the system. For example, the hologram without speckle reduction (Fig. 2(a1)) was captured with a 30.3 ms exposure, whereas activating the LCDS device (Fig. 2(b1)) required increasing the exposure time to 272.6 ms.

Finally, it is worth noting that noisy holograms - Figs. 2(a1) and 2(c1) - reveal not only speckle noise but also intensity parasitic fringes and artifacts likely caused by dust on the illumination objective. Regardless of the source - whether coherent noise or fixed-pattern contamination - the LCDS effectively suppresses these artifacts.

To demonstrate the practical effectiveness of the LCDS speckle reduction, Fig. 3 presents holograms (a1), (b1), reconstructed phase maps (a2), (b2), and amplitude maps (a3), (b3) of *Saccharomyces cerevisiae* acquired without (left column) and with (right column) LCDS-based speckle reduction. In microbiological workflows, monitoring yeast morphology is crucial for assessing culture state, detecting active budding, and identifying possible contamination. The untreated hologram (Fig. 3(a1)) shows that coherent speckle noise obscures fine Gabor fringes, particularly in dense cell clusters, and introduces background fluctuations. With LCDS suppression (Fig. (b1)), speckle artifacts are markedly reduced, yielding smoother backgrounds and higher fringes contrast. Phase reconstructions (Fig. 3(a2), (b2)) clearly reveal the characteristic spherical yeast morphology, while amplitude maps (Fig. 3(a3), (b3)) highlight the improved definition of cell outlines in the denoised case. These results show that LCDS spatial coherence modulation enhances both qualitative visualization and quantitative morphological assessment in dense biological samples.

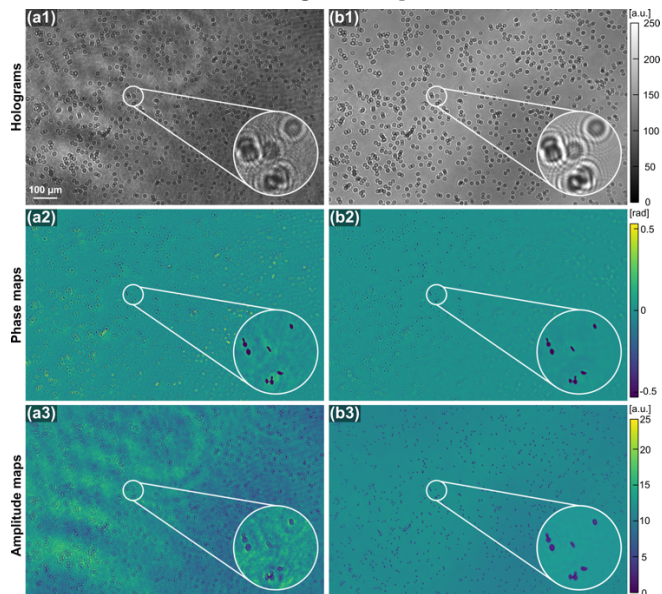

Fig. 3. Holograms and reconstructions of fixed yeast cells imaged with and without LCDS speckle reduction. (a1, b1) Raw holograms acquired without and the LCDS. Corresponding phase (a2, b2) and amplitude (a3, b3) maps reconstructed in the focal plane.

## CONCLUSION

We have demonstrated that LCDS speckle reduction can significantly suppress coherent noise in in-line DHM while preserving - and in some cases enhancing - the system's spatial resolution, all while maintaining simplicity. Integrated into a minimally modified bright-field microscope, the LCDS cell operates without moving parts, providing temporally varying speckle patterns through electrically driven modulation. Experiments on USAF resolution targets and biological samples confirm that the LCDS approach improves the quality of both holograms and reconstructed phase and amplitude maps. Importantly, the system retains its ability to preserve dense Gabor fringes encoding the complex optical field information in the in-line hologram. These results position LCDS speckle reduction as compact, vibration-free alternative to mechanical diffusers, particularly well-suited for quantitative phase imaging in constrained or integrated optical systems.

**Funding.** Research was funded by the Warsaw University of Technology within the Excellence Initiative: Research University (IDUB) programme (XIII Mobility PW) and project no. WPC3/2022/47/INTENCITY/2024 funded by the National Centre for Research and Development (NCBR) under the 3rd Polish-Chinese/Chinese-Polish Joint Research Call (2022). E.W. was supported by National Science Center, Poland (2020/37/B/ST7/03629). N.P.S. gratefully acknowledges the Engineering and Physical Sciences Research Council (EPSRC) UK for financial support through a graduate student scholarship (EP/T517811/1). S.M.M., M. J. B., and S.J.E., gratefully acknowledge financial support from the EPSRC UK through project EP/W022567/1.

**Disclosures.** The authors declare no conflicts of interest.

**Data Availability.** Data underlying the results presented in this paper are not publicly available at this time but may be obtained from the authors upon reasonable request.

**Supplemental Document.** Video_1.